\documentstyle[prd,aps,psfig,twoside]{revtex}
\newcommand{\be}{\begin{equation}}
\newcommand{\ee}{\end{equation}}
\newcommand{\bea}{\begin{eqnarray}}
\newcommand{\eea}{\end{eqnarray}}
\newcommand{\hub}{{\cal H}}

\begin{document}

\begin{flushright}
hep-th/0207065
\end{flushright}

\begin{center}
{\Large\bf  Does curvature-dilaton coupling with Kalb Ramond field lead to an
accelerating Universe ?}
\\[15mm]
Soumitra SenGupta \footnote{E-mail: tpssg@mahendra.iacs.res.in} \\
{\em Department of Theoretical Physics, Indian Association for the
Cultivation of Science,\\ Calcutta - 700 032, India}

Saurabh Sur \footnote{E-mail: saurabh@juphys.ernet.in}\\
{\em Department of Physics, Jadavpur University, Calcutta - 700 032, India}\\[15mm]
\end{center}

\begin{abstract}
In this work we show that  the Universe evolving in a spacetime with
torsion (originated from a second rank antisymmetric Kalb-Ramond field)
and dilaton is free from any big bang singularity and
can have acceleration during the evolution. Both the matter and radiation
dominated era have been considered and the role of the dilaton to explain the
decelerating phase in the earlier epoch has also been discussed.
\end{abstract}

\bigskip
\bigskip

\section{Introduction}

Eversince Einstein-Cartan (EC) theory was proposed, spacetime {\it torsion} has
provided a substantial amount of interest both in gravity and cosmology. Such a theory,
however, is faced with serious difficulties in absence of any confirmed experimental
signature in favour of torsion. There was a revival in the interest for such a theory
after the development of String theory where the massless second-rank antisymmetric
Kalb-Ramond (KR) field $B_{\mu\nu}$ \cite{kr} arising in the heterotic string spectrum
has a natural explanation in the form of spacetime torsion. It has recently been shown
\cite{pmssg} that in the string theoretic scenario, spacetime torsion can in fact be
identified with KR field strength augmented with the Chern-Simons (CS) extension.
Such an extension, which originates on account of cancellation of gauge anomaly in the
context of string theory, plays the crucial role in restoring electromagnetic $U(1)$
gauge invariance which is normally lost in the standard EC framework \cite{pmssg}. This
enables one to study the effect of torsion on the propagation of electromagnetic waves
preserving the gauge symmetry. Investigations have also been carried out in the
context of theories with large extra dimensions, namely those of Arkani
Hamed--Dimopoulos--Dvali (ADD) \cite{add} and Randall--Sundrum (RS) \cite{rs}. An interesting
possibility in such models is that of torsion existing in the bulk together with gravity,
while all the standard model fields are confined to a 3-brane. In order to study the effect
of extra dimensions on the spacetime torsion in the context of RS scenario it has been
shown by Mukhopadhyaya et.al. \cite{bmsenssg} that in a RS type of model where the torsion has the same status
as gravity in the bulk, the effects of massless torsion becomes heavily suppressed on the
standard model brane, thus producing the illusion of a torsionless Universe. This
explains why the detection of torsion has been eluding us for so long.

In recent years, it has already been shown that the presence of KR field in the
background spacetime may lead to various interesting astrophysical/cosmological
phenomena like cosmic optical activity, neutrino helicity flip, parity violation
etc \cite{ssg}. This motivates us to address some of the important problems associated
with the standard Friedmann-Robertson-Walker (FRW) model in the light of KR cosmology.
The most longstanding problem is the existence of so-called `big bang singularity' and
the most recent one concerns the observational evidences from type Ia Supernova data
\cite{riess,perl} showing {\it acceleration} at a late time (may even be at the present
epoch considering the data from the WMAP \cite{wmap}) in course of evolution
of the Universe. A considerable amount of work has
been carried out in this regard in the  past in the framework of FRW cosmology or
in alternative cosmological models \cite{coplidsey}. However, the models obeying
the `cosmological principle', i.e., the large scale homogeneity and isotropy of the
Universe, are of more significance for obvious observational reasons. Although some
bouncing solutions were found in presence of torsion in the past \cite{traut}, the present
work reexamines these scenarios in a homogeneous and isotropic cosmological
model originated from a background spacetime with torsion along with dilaton.

\section{General Formalism}

Following the formalism in \cite{pmssg}, the action for gauge invariant
Einstein-Cartan-Kalb-Ramond coupling in presence of external matter
fields is given by
\be
S ~=~ \int ~d^4 x ~\sqrt{- g} ~\left[ \frac{R(g, T)}{2 \kappa}  ~-~ c_1 H_{\mu\nu
\lambda} H^{\mu\nu\lambda} ~+~ \frac{c_2}{\sqrt{\kappa}} T_{\mu\nu\lambda}
H^{\mu\nu\lambda} ~+~ L_{m} \right]
\ee
where $\kappa ~=~ 8 \pi G/c^4$ is the gravitational coupling constant and
$T_{\mu\nu\lambda}$ is the torsion tensor which is the antisymmetrization of the
affine connection in EC spacetime and is chosen to be antisymmetric in all its indices.
$H_{\mu \nu \lambda}$ is the strength of the KR field $B_{\mu\nu}$ plus the Chern-Simons
(CS) term $\Omega_{cs}$ which keeps the corresponding quantum theory
anomaly-free: ~${\bf H} = d {\bf B} + \Omega_{cs}$.~~$\Omega_{cs}$, however, contains
a suppression factor of the order of Planck mass relative to $d {\bf B}$ and therefore can
safely be dropped from ${\bf H}$ in the present analysis. In the above action
$L_m$ is the external matter Lagrangian density and $c_1$ and $c_2$
are respectively the coupling constants for self-coupling of the three-form ${\bf H}$
and the ${\bf H - T}$ coupling. The Ricci scalar $R(g, T)$ in the torsioned spacetime
is related to the pure Einsteinian scalar curvature $R(g)$ as
\be
R(g, T) ~=~ R(g) ~-~ T_{\mu\nu\lambda} T^{\mu\nu\lambda}.
\ee
The torsion tensor $T_{\mu \nu \lambda}$ being an auxiliary field in the
action (1), satisfies the constraint relation
\be
T_{\mu \nu \lambda} ~=~ \frac{2 c_1}{c_2} \sqrt{\kappa} ~H_{\mu \nu \lambda}
\ee
which implies that the augmented KR field strength 3-tensor acts as the
source of torsion \cite{hehl}.
The resulting action, which has direct correspondence with heterotic string theory,
is then given by
\be
S ~=~ \int d^4 x~ \sqrt{- g} \left( \frac{R(g)}{2 \kappa} ~+~ c_T~ H_{\mu
\nu \lambda} H^{\mu \nu \lambda} ~+~ L_{m} \right)
\ee
where $c_T = c_1 (1 - 2 c_1/c_2^2)$.

Now, in the string scenario the low-energy effective action compactified to four
spacetime dimensions contains alongwith the KR field, the massless scalar dilaton
field $\phi$. The low
energy heterotic string action in the String frame is then expressed as
\be
S ~=~ \int d^4 x~ \sqrt{-g} \left[e^{-\phi} \left(\frac{R(g) + [D\phi]^2}
{2\kappa} ~+~ c_T~ {\bf H}^2\right) ~+~ L_{m}\right]
\ee
where ${\bf H}^2 = H_{\mu \nu \lambda} H^{\mu \nu \lambda}$ and $[D\phi]^2 =
D_{\mu} \phi D^{\mu} \phi$;~$D_{\mu}$~ being the covariant derivative defined in
terms of the usual Christoffel connection. $L_m$ is the Lagrangian density for
external matter fields, for example, the standard cosmological matter (perfect fluid).

Now, as a general convention, specifically in observational studies, it is useful to work
in the Einstein frame which can be obtained from the String frame by making a conformal
transformation $g_{\mu\nu} \rightarrow \tilde{g}_{\mu\nu} = e^{-\phi}  g_{\mu\nu}$.
The resulting 4-dim effective string action is given by
\be
\tilde{S} ~=~ \int d^4 x~ \sqrt{-\tilde{g}} \left[\frac{\tilde{R}(\tilde{g}) - \frac 1 2
[\tilde{D}\phi]^2}{2\kappa}  ~+~ c_T ~e^{- 2\phi}~
\tilde{{\bf H}}^2 ~+~ \tilde{L}_m\right]
\ee
where $\tilde{H}_{\mu \nu \lambda} = H_{\mu \nu \lambda}$ and $\tilde{L}_m =
e^{2 \phi} L_m$, the untilded quantities refer to the String frame.  Dropping the
tildes for simplicity one can express the corresponding field equations as
\bea
R_{\mu \nu} - \frac 1 2 g_{\mu \nu} R &=& \kappa~\left\{T_{\mu \nu}^{(\phi)} ~+~
T_{\mu \nu}^{(m)} ~+~ e^{-2 \phi}~T_{\mu \nu}^{(H)} \right\} \\
D_{\mu} \left(e^{- 2\phi}~ H^{\mu \nu \lambda}\right)  &=& 0 \\
D_{\mu} \phi^{\mu} ~=~ 4 \kappa~(c_T~e^{-2 \phi}~{\bf H}^2 &-& T_{\mu}^{~\mu~(m)})
\eea
where  $\phi_{\mu} \equiv \partial_{\mu} \phi$.~$T_{\mu \nu}^{(m)}$ is the
energy-momentum tensor corresponding to the background matter distribution
which for our cosmological model is taken to be a perfect fluid, and $T_{\mu
\nu}^{(\phi)}$ and $T_{\mu \nu}^{(H)}$ are the analogous contributions due to the
dilaton and the KR field:
\bea
T_{\mu \nu}^{(\phi)} &=& \frac 1 {2 \kappa} (\phi_{\mu}
\phi_{\nu} - \frac 1 2 g_{\mu \nu} \phi_{\alpha} \phi^{\alpha}) \\
T_{\mu \nu}^{(m)} &=& \left[ (\rho c^2 + p) u_{\mu} u_{\nu} ~-~ p ~g_{\mu \nu} \right] \\
T_{\mu \nu}^{(H)} &=& - ~6~ c_T~ ( 3 H_{\alpha \beta \mu}
H^{\alpha \beta}_{~~~\nu} ~-~ \frac 1 2 g_{\mu \nu} H_{\alpha \beta \gamma}
H^{\alpha \beta \gamma} )
\eea
$p$ and $\rho$ being the pressure and mass-density respectively, and $u_{\mu} =
(1, 0, 0, 0)$ is the hypersurface-orthogonal four-velocity vector.

Now, with the CS term neglected, the three form ${\bf H}$ being equal to the exterior
derivative of the two form ${\bf B}$, i.e., ${\bf H} = d {\bf B}$, one can verify the
Bianchi identity in 4-dim
\be
d {\bf H} ~=~ \epsilon^{\mu\nu\lambda\beta} \partial_{\beta} \partial_{[\mu}
B_{\nu\lambda]} ~=~ 0
\ee
which alongwith Eq.(8) leads to the well-known duality relationship in string
theory between the three form ${\bf H}$ and the pseudoscalar ({\it axion}) $\xi$:
\be
H^{\mu \nu \lambda} ~=~ \epsilon^{\mu\nu\lambda\beta} ~e^{2 \phi}~ \xi_{\beta}.
\ee
where $\xi_{\mu} \equiv \partial_{\mu} \xi$ and $\xi$ satisfies the wave equation
\be
D_{\mu} (e^{2 \phi}~ \xi^{\mu}) = 0.
\ee
In our subsequent analysis we shall be using the axion $\xi$ frequently
rather than the KR field strength ${\bf H}$ using the above duality.

In order to have a large-scale homogeneous and isotropic cosmological model, we
consider the standard Robertson-Walker metric structure:
\be
ds^2 ~=~ c^2 dt^2 ~-~ a^2(t)~ \left[ \frac{dr^2}{1 - k r^2} ~+~ r^2 ( d\theta^2 ~+~
\sin^2 \theta d\phi^2 ) \right]
\ee
$a(t)$ being the scale-factor and $k$ is the curvature. It can now easily be
shown that for this type of metric the consistency of the field equations
immediately demands that both the axion $\xi$ and the dilaton $\phi$ depend
on time only. Eq.(15) then yields
\be
\xi_0 ~\equiv~ \frac 1 c \frac d {dt} \xi ~=~ \frac{\alpha}{a^3 (t)} e^{-2 \phi}
\ee
$\alpha$ being a parameter determining the strength of the axion. The effective
Kalb-Ramond energy density is then given by
\be
\rho_{KR} ~\equiv~ \frac 1 {c^2} T_0^{0~(KR)} ~=~ - \frac{6~c_T~\alpha^2}{c^2 a^6(t)}.
\ee
The sign of $c_T$ determines whether the KR energy density is positive or negative.

The field equations now reduce to
\bea
\frac{\dot{a}^2}{a^2} ~+~ \frac{k c^2}{a^2} &=& \frac{\dot{\phi}^2}{12} ~+~ \frac{8
\pi G}{3 c^2}~\left( \rho c^2 ~-~ \frac{6~c_T~\alpha^2~e^{- 2 \phi}}{a^6}  \right)\\
\frac{\ddot{a}} a &=& -~ \frac{\dot{\phi}^2} 6 ~-~ \frac{8 \pi G} {3 c^2}~\left( \frac{\rho c^2
+ 3 p} 2 ~-~ \frac{6~c_T~\alpha^2~e^{- 2 \phi}}{a^6} \right) \\
\frac d {dt} (a^3 \dot{\phi}) &=& -~ \frac{32 \pi G}{c^2} ~\left\{ (\rho c^2 - 3 p)~a^3 ~+~
\frac{6~c_T~\alpha^2~e^{- 2 \phi}}{a^3}\right\}
\eea
where overhead dot stands for derivative with respect to time. For a general
equation of state $p = \omega \rho c^2$ we rewrite the field equations as
\bea
\frac{\dot{b}^2}{b^2} ~+~ \frac{\chi}{b^2} &=& \frac{\dot{\phi}^2}{12}
~+~ \frac{\gamma}3~\left(\zeta ~-~ \frac{\sigma}{b^6} e^{-2\phi}\right) \\
\frac{\ddot{b}}{b} &=& -~ \frac{\dot{\phi}^2}6 ~-~ \frac{\gamma}3~\left(\frac{1 + 3
\omega}2 \zeta ~-~ \frac{2 \sigma}{b^6} e^{-2\phi} \right) \\
\frac d{dt} (b^3 \dot{\phi}) &=&  -~ 4 \gamma ~\left\{ (1 - 3 \omega) \zeta b^3
~+~ \frac{\sigma}{b^3}~e^{-2\phi} \right\}
\eea
where, for convenience, we have used dimensionless quantities:
\be
b = \frac a{a_0}~,~~~\zeta = \frac{\rho}{\rho_0}~,~~~\chi = \frac{k c^2}{a_0^2}~,~~~
\gamma = 8 \pi G \rho_0~,~~~\sigma = \frac{6~c_T~\alpha^2~a_0^{-6}}{\rho_0~c^2}.
\ee
The subscript $0$ generally stands for the values of the quantities at the present
epoch $t_0$. Recalling Eq.(18), one finds the quantity $\sigma$ defined above gives
a relative measure of the present values of KR field energy density and the energy
density of the Universe. From the above field equations it is easy to show that $\zeta$
satisfies the energy-momentum conservation relation
\be
\frac d {db} (\zeta b^3) ~=~ -~ 3 \omega \zeta b^3 \left\{ \frac{3 \omega} b ~-~
2 (1 - 3 \omega)~ \frac{d\phi}{db} \right\}
\ee
with solution
\be
\zeta ~=~ \frac{e^{2 (1 - 3 \omega) \phi}}{b^{3 (1 + \omega)}}.
\ee
Now, in the development of a cosmological model the quantity $\sigma$ may be positive
or negative depending on the sign of the torsion-KR coupling parameter $c_T$. This implies
that the effective KR energy density $\rho_{KR}$ can be negative as well as positive. In
the case of low energy string effective action the term quadratic in the three-form ${\bf H}$
appears with a negative ($= - 1/12$) sign which ensures positive energy density for the
KR field. Therefore, if one persists in considering
the spacetime torsion to be identified with the KR field strength then
$\sigma$ can only take negative values.
On the contrary, a repulsive (anti-gravitating) character of torsion is well known
in the context of cosmologies based on Einstein-Cartan theory. Bouncing cosmological
solutions have already been shown to exist in Bianchi type torsioned spacetimes \cite{traut}.
Thus one is led to the fact that even when torsion is generated by the hitherto attractive
KR field in string background geometries of Riemannian type, the repulsive character of torsion
may still be restored via the nature of its coupling with the KR field within the minimal
coupling prescription when $\sigma > 0$ (or, in other words when the effective KR energy
density is negative). This may well be treated as a departure from the usual
string theoretic picture and can be looked upon as an alternative theory of gravity.
However, regardless of the sign of $\sigma$ we show here the existence of a theory
of gravity with torsion where the background metric continues to be homogeneous
and isotropic and is described by a structure similar to Robertson-Walker spacetime.
We seek specific
solutions of the above field equations in a scenario where $\sigma$ can be positive as well
as negative, separately for the present matter-dominated (dust) Universe and for the
early radiation-dominated Universe.

\section{Matter-dominated Universe}

Presently, in the Universe ~$p \ll \rho c^2$,~i.e., effectively ~$\omega = 0$, and
hence $\zeta = e^{2 \phi}/b^3$. The dilatonic evolution rate $\dot{\phi}$ as well as
the Hubble parameter $\hub$ and the deceleration parameter $q$ can be obtained from the
field equations (25) - (27):
\bea
\dot{\phi}^2 &=& \frac{4 \gamma}{b^6}~\left\{ \sigma e^{- 2 \phi} ~-~ b^3 e^{2 \phi}
~+~ 3 f(b) ~+~ \Lambda_m \right\}~;~~~~ f(b) = \int e^{2 \phi} b^2 db \\
\hub &\equiv& \frac{\dot{b}}b ~=~ \left[ -~ \frac{\chi}{b^2} ~+~ \frac{\gamma}{3 b^6}~
\{3 f(b) ~+~ \Lambda_m\} \right]^{1/2} \\
q &\equiv& -~\frac 1 {\hub^2}~\frac{\ddot{b}} b ~=~ \frac{\gamma}{3 \hub^2 b^6}~
\left\{ 6 f(b) ~+~ 2 \Lambda_m ~-~ \frac 3 2 b^3 e^{2 \phi} \right\}.
\eea
Imposing the limiting conditions:~ $\phi = \phi_0 ,~ \dot{\phi}^2 = \lambda_0 ,~ \hub =
\hub_0$ at the present epoch $t = t_0 ~(b = 1)$ whence $f(b) = f(1)$, the constant
$\gamma$ as well as the integration constant $\Lambda_m$ turn out to be
\bea
\gamma &=& \frac{12 (\hub_0^2 + \chi) ~-~ \lambda_0}{4 (e^{2 \phi_0} ~-~ \sigma e^{- 2
\phi_0})}~,
~~~~~ \sigma \neq e^{4 \phi_0} \nonumber \\
\Lambda_m &=& \frac 3 {\gamma}~\left\{ \hub_0^2 ~+~ \chi ~-~ \gamma f(1) \right\}
\eea
The Hubble parameter and the deceleration parameter reduce to
\bea
\hub &=& \frac 1 {b^3}~ \left[ \hub_0^2 ~+~ \gamma~\{ f(b) - f(1) \} ~+~ \chi~
(1 - b^4) \right]^{1/2} \\
q &=& 2 ~-~ \frac{ \gamma e^{2 \phi} ~-~ 4 \chi b }{2 \hub^2 b^3}
\eea
As can be seen from the above equations that an accelerating Universe ($\ddot{b} > 0 ,~
q < 0$) is only possible
when $\gamma > 4 (\hub^2 b^3 + \chi b) e^{- 2 \phi}$. In late stages of the evolution
of the Universe this implies that for $\sigma < 0$ i.e., $\rho_{KR} > 0$, acceleration is
never possible. This can be clearly be observed from Eq.(23) which
for the matter-dominated model takes the form
\be
\frac{\ddot{b}}{b} ~=~ -~ \frac{\dot{\phi}^2}6 ~-~ \frac{\gamma}3~
\left(\frac{e^{2 \phi}}{2 b^3} ~-~ \frac{2 \sigma e^{-2\phi}}{b^6} \right).
\ee
As $\dot{\phi}^2$ is known to be positive in the string theoretic
scenario, an accelerating Universe at any of stage of its evolution is not
possible in pure stringy cosmological models where $\sigma < 0$ as pointed out
earlier. However, if the nature of the KR-torsion coupling be such that $\sigma > 0$
then torsion would exhibit an anti-gravitating character and in such case an
accelerating cosmological model is indeed possible whenever the KR field is
sufficiently large to overcome the effect of the dilaton
(considered to be of positive energy density) and ordinary gravitating matter.
We now make a more critical examination of the scenario resorting to a rather
typical dilatonic evolution.

\vskip .2in
\noindent
{\bf Frozen dilaton :}
\vskip .1in

As we are primarily interested in the effect of KR field on the standard Robertson-Walker
cosmology especially in late times, we consider dilatonic evolution rate gradually gets
damped from its value in the early stages and finally the dilaton `freezes out' to a constant
vacuum expectation value (vev) $= \phi_0$ at an epoch $t_f < t_0$. In such a scenario one
has the limiting conditions:

(i)~~$\phi = \phi_0$ for the entire period $t_f \le t \le t_0$ i.e.,
for $b_f \le b \le 1$ where $b_f = b \mid_{t = t_f}$;

(ii)~~$f(b) = e^{2 \phi_0} b^3/3$ and $\dot{\phi} = 0$ for $t_f < t \le t_0$ ($b_f < b \le 1$).

\noindent
These imply $f(1) = e^{2 \phi_0}/3$ and $\lambda_0 \equiv \dot{\phi}^2 \mid_{t = t_0} = 0$.
The Hubble parameter and the deceleration parameter take the forms:
\bea
\hub &=& \left[ -~ \frac{\chi}{b^2} ~+~ \frac{\gamma}{3 b^6}~\{ 3 f(b) ~-~ \sigma
e^{- 2 \phi_0}\} \right]^{1/2}~;~~~~~~
\gamma ~\equiv~ 8 \pi G \rho_0 ~=~ \frac{3 (\hub_0^2 + \chi) ~e^{2 \phi_0}}
{e^{4 \phi_0} ~-~ \sigma}~;~~~~ (\sigma \neq e^{4 \phi_0}) \\
q &=& 2 ~-~ \frac 3 2~\left[ \frac{b^3 e^{2 (\phi + \phi_0)} (\hub_0^2 ~+~ \chi) ~-~
\frac 4 3~\chi b^4 (e^{4 \phi_0} - \sigma)} {\{3 f(b) e^{2 \phi_0} - \sigma\} (\hub_0^2 ~+~
\chi) ~-~ \chi b^4 (e^{4 \phi_0} - \sigma)} \right]
\eea
In addition one obtains from Eq.(28) the expression for the dilatonic evolution
rate ($d\phi/db$):
\bea
\left(\frac{d\phi}{db}\right)^2 &=& \frac{12 \hub_0^2}{b^8 \hub^2}~\left\{ 1 ~+~ \frac
{3 f(b) ~-~ b^3 e^{2\phi}}{e^{2\phi_0} ~-~ \sigma e^{- 2\phi_0}}\right\}~;~~~~~~
f(b) ~=~ \int^b ~e^{2 \phi (b)}~ b^2 db ~;~~~~~~~~~~~ t \le t_f ~~(b \le b_f) \nonumber\\
\left(\frac{d\phi}{db}\right)^2 &=& 0 ~;~~~~~~~~~~~~~~~~~~~~~~~~~~~~~~~~~~~~~~~~~~~~~~
f(b) ~=~ \frac{e^{2 \phi_0}~ b^3} 3 ~;~~~~~~~~~~
 t_f < t \le t_0 ~~(b_f < b \le 1) .
\eea
As a simplification we consider only the spatially flat matter-dominated model setting
the curvature $\chi = k c^2/a_0^2$ equal to zero. This is fairly justified considering
the fact that in the standard Friedman-Robertson-Walker (FRW) framework the
matter-dominated
Universe is known to have negligible curvature, one expects that the effect of curvature
on our results for a spatially flat model involving the dilaton and the axion may not be
of much significance.

\vskip .15in
\noindent
{\bf I.~ $b > b_f$ era:}
\vskip .1in

The expressions for $\hub$ and $q$ can be given for $\chi = 0$ and $b > b_f$ as
\bea
\hub &\equiv& \frac{\dot{b}}b ~=~ \frac 1 {b^3}~\sqrt{\frac{\gamma} 3~
( b^3 e^{2 \phi_0} ~-~ \sigma e^{- 2 \phi_0} )}~;~~~~~~~~
\gamma ~\equiv~ 8 \pi G \rho_0 ~=~ \frac{3 \hub_0^2 ~e^{2 \phi_0}}
{e^{4 \phi_0} ~-~ \sigma}  \\
q &=& 2 ~-~ \frac 3 2~\left(\frac{b^3 e^{4 \phi_0}}{b^3 e^{4 \phi_0} ~-~ \sigma}\right) .
\eea
For $\sigma > 0$, the constant $\gamma = 8 \pi G \rho_0$ being positive definite,
one obtains an upper bound on $\sigma$ :~ $\sigma < e^{4 \phi_0}$.~ This maximum
limit on $\sigma$ is further reduced on taking into account the fact that $\hub^2$
is positive always for all $b > b_f$, which implies ~ $\sigma < b_f^3 e^{4 \phi_0}$.~
Moreover, to have an accelerating Universe ($q < 0$) at any value $b_{accl}~ (> b_f)$
the value of $\sigma$ should be limited by ~$\sigma > (1/4) b_{accl}^3 e^{4 \phi_0}$.
Consistency of these two bounds on $\sigma$ then demands ~$b_{accl}^3 > 4 b_f^3$.
The ultimate bounds posed on $\sigma$ and $b_f$ in the context of present-day accelerating
Universe ($q_0 < 0$) therefore turn out to be
\be
\frac 1 4~e^{4 \phi_0} ~<~ \sigma ~<~ b_f^3~e^{4 \phi_0}~;~~~~~~~~~~~ b_f ~>~ 4^{- 1/3}.
\ee

\noindent
For $\sigma < 0$, $\gamma$ and $\hub^2$ being positive by construction,
no limit is imposed on
$\sigma$. Moreover, from Eq.(39) $q$ is always positive and greater than $0.5$ (i.e.,
it's value in spatially flat FRW model) which in turn signifies absence of any
accelerating phase of the Universe all the way through the post-frozen dilaton era $b > b_f$.

\noindent
For $\sigma = 0$, however, one recovers the FRW result $q = 0.5 = q_{FRW}$.

Regardless of the sign of $\sigma$, the functional form of the scale factor $b(t)$ can
be obtained straightaway on solving
Eq.(38). Given the boundary conditions $b = b_f, ~\phi = \phi_0$ at $t = t_f$ and
$b = 1, ~\phi = \phi_0, ~\hub = \hub_0$ at $t = t_0$ the solution is expressed as
\be
b (t) ~=~ \left[ b_f^3 ~+~ 3 \hub_0~\left(\frac{b_f^3 e^{4 \phi_0} ~-~
\sigma}{e^{4 \phi_0} ~-~ \sigma}\right)^{1/2}(t - t_f) ~+~ \frac {9 \hub_0^2} 4 ~
\left(\frac{e^{4 \phi_0}}{e^{4 \phi_0} ~-~ \sigma}\right)~(t - t_f) \right]^{1/3} .
\ee
In the limit $\sigma \rightarrow 0$, $b \rightarrow \{ b_f^{3/2} + (3 \hub_0/2) (t - t_f)
\}^{2/3}$ which shows a similar behaviour as the spatially flat matter-dominated FRW
Universe ($b_{FRW} \sim t^{2/3}$) especially when the dilaton freezes at a very
early epoch $t_f \ll t_0~ (b_f \ll 1)$.

The post-frozen dilaton epoch time lapse is given by
\be
T_f ~\equiv~ (t_0 - t_f) ~=~ \frac 2 {3 \hub_0} ~(1 ~-~ \sigma e^{- 4 \phi_0}) ~\left\{
1 ~-~ \left( \frac{b_f^3 e^{4 \phi_0} ~-~ \sigma}{e^{4 \phi_0} ~-~ \sigma}\right)^{1/2}
\right\}
\ee
which has the limiting form $T_f = 2 (1 - b_f^{3/2})/3 \hub_0$ as $\sigma \rightarrow 0$,
and is almost the same as the age of the matter-dominated Universe in the FRW framework
if the dilaton freezes in a very early epoch ($b_f << 1$).

The variations of the scale factor $b$ and the deceleration parameter $q$ with
time in the post-frozen dilaton era has been depicted in Fig.1 for three
characteristic values of $\sigma~(= - 0.35, 0, + 0.35)$. The time scale is as usual
taken to be the inverse Hubble constant $\hub_0$.
Although the value of $b_f$ may depend on $\sigma$ as well the nature of the evolution
of the Universe in the pre-frozen dilaton era, we have for simplicity chosen a typical
parametric value ($b_f = 0.75$) for all the three values of $\sigma$. The nature of the
curves are however not altered by this simplification.
\begin{figure}
\psfig{figure=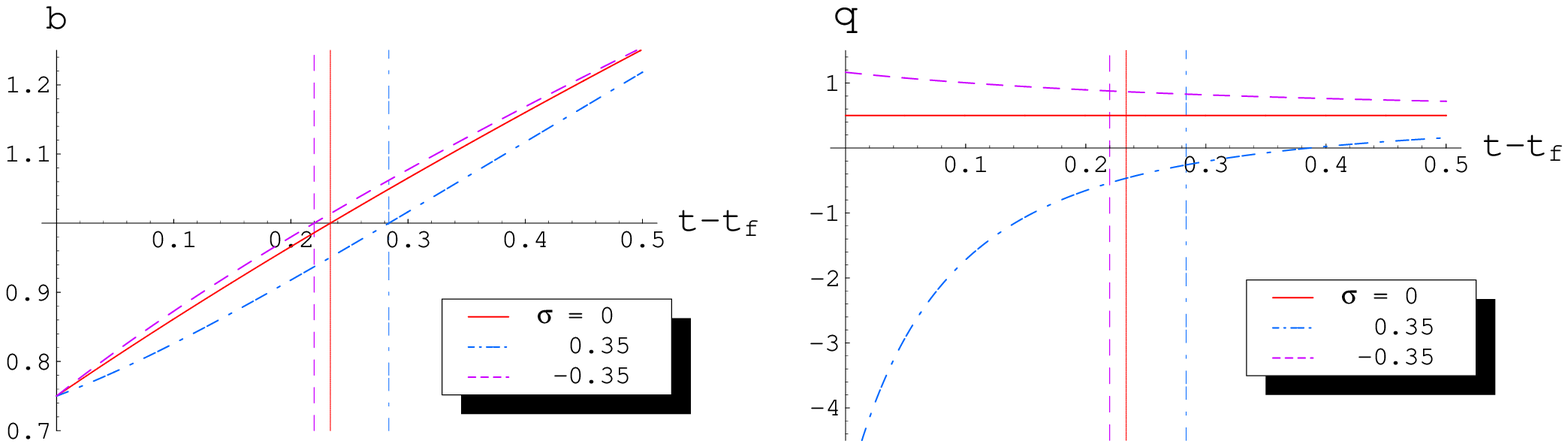,width=7in,height=2.25in}
\caption{\it $b - t$ and $q - t$ plots for spatially flat matter-dominated Universe
in the post-frozen dilaton era $b > b_f$ for three parametric values of $\sigma$,
+0.35, 0, -0.35. The time-scale is, as usual, the inverse Hubble constant and $b_f$ is
characteristically chosen to be equal to 0.75. The vertical grids denote the present
time-slice $t_0 = const.$ for various $\sigma$ while the present scale factor is given
by $b=1$.}
\end{figure}

It should also be noted here that considering the late time accelerating expansion
of the Universe as confirmed by the Supernova Ia data \cite{riess,perl} and the recent
WMAP results we focus primarily on a model with $\sigma > 0$ (i.e., negative effective
energy density of the KR field) where one can find such acceleration as shown above.

\vskip .15in
\noindent
{\bf II.~ $b \le b_f$ era:}
\vskip .1in

The expressions for $\hub$ and $q$ for $\chi = 0$ and $b \le b_f$ are given by
\bea
\hub &\equiv& \frac{\dot{b}}b ~=~ \frac 1 {b^3}~\sqrt{\frac{\gamma} 3~
\{ 3 f(b) ~-~ \sigma e^{- 2 \phi_0} \} }~;~~~~~~~~
\gamma ~\equiv~ 8 \pi G \rho_0 ~=~ \frac{3 \hub_0^2 ~e^{2 \phi_0}}
{e^{4 \phi_0} ~-~ \sigma}  \\
q &=& 2 ~-~ \frac 3 2~\left\{\frac{b^3 e^{2 (\phi + \phi_0)}}{3 e^{2 \phi_0} f(b) ~-~
\sigma}\right\}
\eea
Since $\gamma > 0$ we again find that if $\sigma$ is positive then it should
have a value less than $e^{4 \phi_0}$. The fact that $\hub^2$ is positive for
all values of $b$ then puts a further stringent upper bound on $\sigma$:~ $\sigma
< 3 e^{2 \phi_0} f_{min}$, where $f_{min}$ is the minimum value of the integral $f(b)
= \int e^{2 \phi} b^2 db$, i.e., the value of $f$ when $b = b_m$ -- the root
of the equation ~$e^{2 \phi (b_m)} b_m^2 = 0$.

The expression for evolution rate of the dilaton [Eq.(37)] in the present
circumstances reduces to
\be
\left(\frac{d\phi}{db}\right)^2 ~=~ \frac{12}{b^2}~\left( 1 ~-~ \frac{b^3 e^{2 \phi}
~-~ \sigma e^{- 2 \phi}}{3 f(b) ~-~ \sigma e^{- 2 \phi_0}} \right).
\ee
Using this and the definition of $f(b)$ one finally obtains
\bea
\hub^2 &=& \frac{\gamma e^{2 \phi}}{3 b^3}~\left( \frac{1 ~-~ \frac{\sigma
e^{-4 \phi}}{b^3}}{1 ~-~ \frac{b^2 \phi'^2}{12}} \right)~;~~~~~~~~~~~~~~
\gamma ~=~ \frac{3 \hub_0^2 ~e^{2 \phi_0}}{e^{4 \phi_0} ~-~ \sigma}  \\
q &=& 2 ~-~ \frac 3 2~ \left( \frac{1 ~-~ \frac{b^2 \phi'^2}{12}}{1 ~-~ \frac{\sigma
e^{-4 \phi}}{b^3}} \right)
\eea
where $\phi' = d\phi/db$.

As for the reasons mentioned earlier in the context of an accelerating Universe,
we now concentrate on the case where $\sigma > 0$ for which one can anticipate
such acceleration. Moreover, some renewed study of the Supernova Ia data provides
evidence of not only a late-time accelerating Universe but a decelerating Universe
in the remote past as well \cite{tr}. This is particularly relevant in order to explain
the structure formation in the early Universe. To understand such a situation
in the present model with $\sigma > 0$ we refer to the above expressions for
the Hubble parameter and the deceleration parameter. The dilaton $\phi$ is
supposed to fall with time (i.e., with $b$ since for an expanding Universe $b$
always increases with time), from a high value in the early Universe until it gets
frozen to $\phi_0$ at $b = b_f$ while $\sigma$ is universally a constant.
The $\phi'^2$ term in Eq.(47) shows that the dilaton only has a positive
contribution to the deceleration parameter $q$ contrary to that of $\sigma$
which gives a negative contribution by reducing the denominator of the second
term on the right of Eq.(47). However, in the early Universe, $\phi$ being larger
the quantity involving $\sigma$ in Eq.(47) is more suppressed by the factor
$e^{4 \phi}$. Therefore, as time progresses the gradual reduction of the dilatonic
effect enables the KR field axion to dominate and this may produce an overall
negative contribution to $q$ at late stages. It should also be noted here that
one can explicitly check that the dilatonic evolution equation (45) in the 
pre-freezeout era is satisfied identically for any functional form of $\phi$ only
under proper consideration of the limiting constraints. However, to make at
least a qualitative assessment of the whole scenario so that the functional
form of $q$ for $b \le b_f$ exactly matches with that given in Eq.(39) for
$b > b_f$ at the freeze-out epoch $t = t_f~(b = b_f)$, we consider a typical
power law type dilatonic fall-off
\be
\phi ~=~ \phi_0 + \left( \frac{b_f} b ~-~ 1 \right)^n
\ee
where $n$ is a positive index. The expressions for $\hub$ and $q$ turn out to be
\bea
\hub^2 &=& \frac{\gamma}{3 b^3}~\exp\left[2 \phi_0 ~+~ 2 \left(\frac{b_f} b ~-~ 1
\right)^n\right]~\left\{  \frac{b^3 ~-~ \sigma ~\exp \left[- 4 \phi_0 ~-~ 4 \left(\frac{b_f} b
~-~ 1\right)^n\right]}{b^3 ~-~ \frac{n^2}{12} ~b~ b_f^2 \left(\frac{b_f} b ~-~ 1\right)^{2
(n - 1)}}\right\} ~;~~~~~
\gamma = \frac{3 \hub_0^2 ~e^{2 \phi_0}}{e^{4 \phi_0} - \sigma} \\
q &=& 2 ~-~ \frac 3 2~ \left\{ \frac{b^3 ~-~ \frac{n^2}{12} ~b~ b_f^2 \left(\frac{b_f}
b ~-~ 1\right)^{2(n - 1)}}
{b^3 ~-~ \sigma ~\exp \left[- 4 \phi_0 ~-~ 4 \left(\frac{b_f} b ~-~ 1\right)^n\right]}
\right\}.
\eea
The nature of $\hub$ vs $b$ and $q$ vs $b$ curve is shown in Fig.2 with parametrically
chosen values of the constants : ~$b_f = 0.75, ~ \phi_0 = 0.01, ~ \sigma = 0.35, ~ n = 1.25$.

\begin{figure}
\psfig{figure=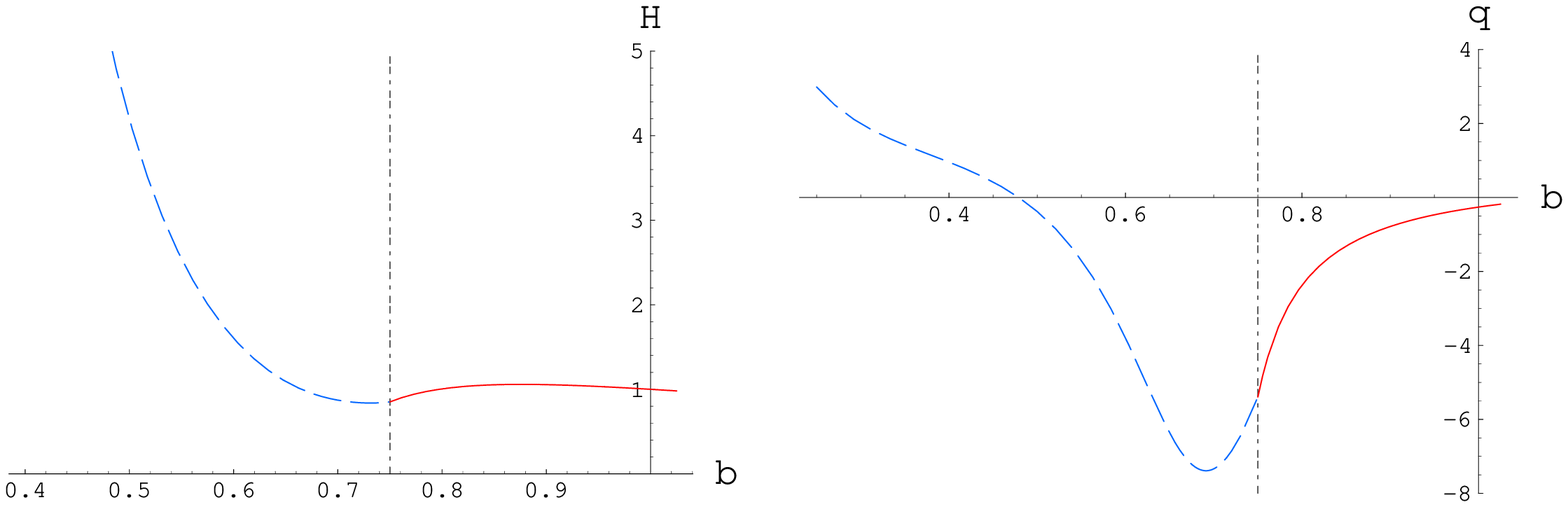,width=7in,height=3in}
\caption{\it $\hub - t$ and $q - t$ plots for spatially flat matter-dominated
Universe for a positive parametric value of $\sigma$ (= 0.35). The time-scale is the
inverse Hubble constant. The vertical grid gives the value $b_f$ of  the scale-factor
when the dilaton freezes out to a vev $\phi_0$. The values of $b_f$ and $\phi_0$
are characteristically chosen to be equal to 0.75 and 0.01 respectively. The present
epoch is given by $b = 1$.}
\end{figure}
\bigskip

\section{Radiation-dominated Universe}

Extrapolation of  the matter-dominated model to a very early epoch $t \ll t_0$ to
study the effects of the dilaton and the axion may not be very reliable because of
the radiation dominance at the early ages. In a radiation-dominated model we have
$p = \frac 1 3 \rho c^2$, i.e., $\omega = 1/3$, and Eq.(27) yields $\zeta = 1/b^4$.
We obtain from the field equations (22) - (24) the expressions for the dilatonic
evolution rate $\dot{\phi} \equiv d\phi/dt$ as well the expressions for the Hubble
parameter $\hub$ and the deceleration parameter $q$:
\bea
\dot{\phi}^2 &=& \frac{\gamma (4 \sigma e^{- 2 \phi} ~+~ \Lambda_r)}{b^6} \\
\hub &\equiv& \frac{\dot{b}} b ~=~ \left[ -~ \frac{\chi}{b^2} ~+~ \frac{\gamma}{12 b^6}
(4 b^2 ~+~ \Lambda_r) \right]^{1/2} \\
q &\equiv& -~ \frac 1 {\hub^2}~\frac{\ddot{b}} b ~=~ \frac{\gamma}{6 b^6 \hub^2}~
(2 b^2 ~+~ \Lambda_r)
\eea
where the parameter $\sigma$ is now the ratio of the KR field energy density and
the present radiation density $\rho_0^{(r)}$ of the Universe and $\gamma = 8 \pi G
\rho_0^{(r)}$. However, since the Universe is presently not radiation-dominated one
cannot determine the integration constant $\Lambda_r$ in terms of the present
values of the physical parameters like the Hubble constant, etc. We, instead, take
the initial values $b = b_i ,~ \phi = \phi_i ,~ \dot{\phi}^2 = \lambda_i$ at the
origin of time $t = t_i$. In terms of these initial values $\Lambda_r$ is given by
\be
\Lambda_r ~=~ \frac{\lambda_i b_i^6}{\gamma} ~-~ 4 \sigma e^{- 2 \phi_i}
\ee
Once again we find from Eqs.(53) and (54) that an accelerated expansion of the
Universe at any phase during its evolution is not possible when $\sigma < 0$, i.e.,
the KR field energy is positive --- the case in usual string background geometries
without torsion. However, our point of interest here is in a KR field induced
torsioned background where a positive value of $\sigma$ is possible. Only in
such case one can have $\Lambda_r$ negative whenever the condition $\sigma >
(4 \gamma)^{-1} \lambda_i b_i^6~ e^{2 \phi_i}$ is satisfied and the Universe can
have an accelerating phase that sets in at some value of $t = t_{accl}$ whence,
$b = b_{accl}$ and $b_{accl}^2 < 2 |\Lambda_r|$, i.e.,
\be
\sigma ~>~ \frac{e^{2 \phi_i}} 4 ~\left( \frac{\lambda_i b_i^6}{\gamma}  ~+~
b_{accl}^2 \right).
\ee
However, for an expanding Universe $b$ being an increasing function of time,
it is expected that the above inequality may break and such an accelerating phase
in the early radiation-dominated era may not last long when $b$ gets much larger
than $b_{accl}$.

For zero curvature $\chi$ (spatially flat Universe) the field equation (52) can be solved
exactly. We write down the solution taking $\Lambda_r$ to be negative (which is
only possible for $\sigma > 0$) in a parametric form as
\bea
b(\eta) &=& \sqrt{\frac{\gamma} 3}~ \left( \eta^2 ~+~ x \right)^{1/2} \\
t(\eta) &=& \sqrt{\frac{\gamma}{12}} \left\{ \eta \sqrt{\eta^2 + x} ~+~ x~ \ln \left(
\frac{\eta ~+~ \sqrt{\eta^2 + x}}{\sqrt{x}} \right) \right\}
\eea
where $x = - 3 (4 \gamma)^{-1}~\Lambda_r$.~ One can explicitly check that the
above solution assumes the FRW form $b \sim t^{1/2}$ if we ignore the effects
of the axion and the dilaton, i.e., $\sigma \rightarrow 0 ,~ \lambda_i =
\dot{\phi}^2|_{t_i} \rightarrow 0$ which implies that $\Lambda_r$ vanishes
identically. In fact, the FRW solution is always recovered in the limit $\Lambda_r
\rightarrow 0$, which is possible even for non-vanishing contributions from
the dilaton and the axion whenever $\lambda_i b_i^6~e^{2 \phi_i} \rightarrow
4 \gamma \sigma$. For non-vanishing $\Lambda_r$ (taken to be negative which implies
$\sigma > 0$), the above solution shows that $b$ remains non-vanishing for
all values of $t$. On back-tracing this spatially-flat radiation-dominated model
to the origin of time $t = t_i, ~\eta = \eta_i$, where both $t_i$ and $\eta_i$
can be set to be equal to zero in view of the limiting FRW case, one finds that
the scale factor $b$ has a non-zero value $(1/2) \sqrt{|\Lambda_r|}$ and hence the
energy density $\zeta = 1/b^4$ is finite at that epoch. This implies the removal
of the so-called `big bang singularity' (that features in FRW cosmology) in a torsioned
spacetime.

The deceleration parameter $q$ have been calculated using the above solution
and the nature of $b - t$ and $q - t$ plots are depicted in Fig.3 for characteristic
values $\sigma = 0.125 ,~ \phi_i = 1 ,~ \lambda_i = 1$ of the physical parameters.
\begin{figure}
\psfig{figure=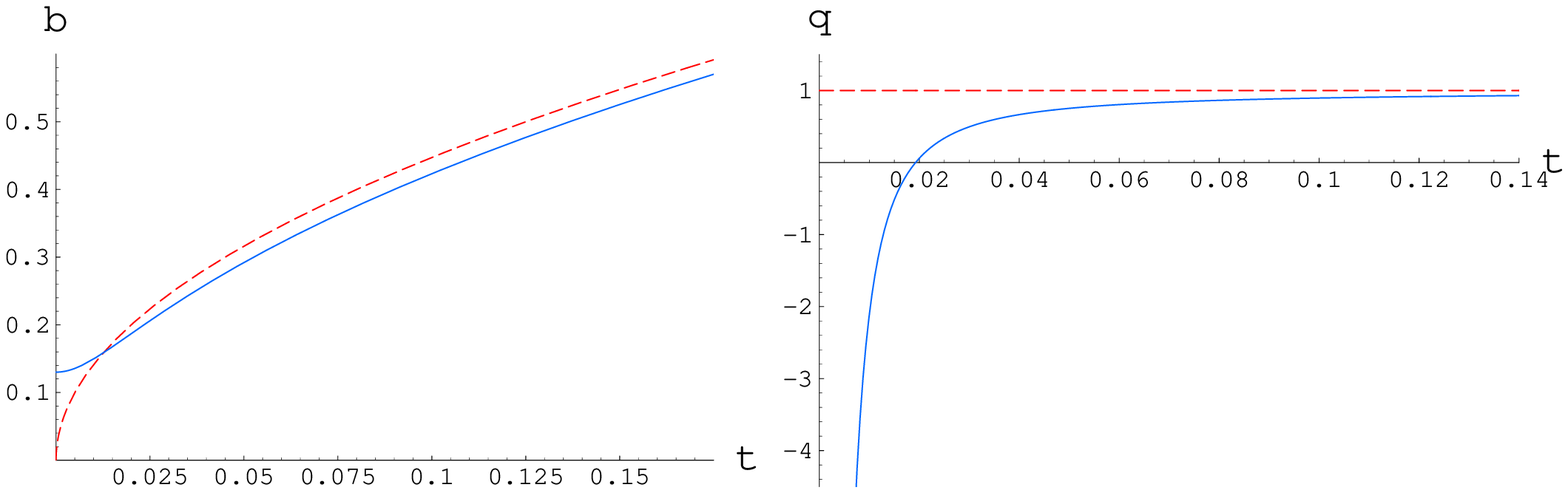,width=7in,height=2.25in}
\caption{\it $b - t$ and $q - t$ plots for spatially-flat radiation-dominated Universe.
The broken line is for FRW model ($\sigma = 0$) while the solid line is for a
model involving the axion and the dilaton ($\sigma = 0.125 ,~ \phi_i = 1 ,~
\lambda_i = 1 $, parametrically chosen).}
\end{figure}

\bigskip
For a spatially non-flat radiation-dominated Universe in presence of the dialton
and the KR field, parametric
solutions of Eq.(52) can be given respectively for closed ($\chi > 0$) and open ($\chi < 0$)
models as
\bea
b(\eta) &=& \frac{\gamma}{6 \chi} ~\left[1 ~-~ \sqrt{1 - \frac{3 \chi \Lambda_r
e^{2 \phi_i}}{2\gamma}}~\cos (2 \eta \sqrt{\chi}) \right]^{1/2}~~~~~~~~~~~~(\chi > 0)\\
b(\eta) &=& \frac{\gamma}{6 |\chi|} ~\left[\sqrt{1 + \frac{3 |\chi| \Lambda_r
e^{2 \phi_i}}{2\gamma}}~\cosh (2 \eta \sqrt{|\chi|}) ~-~ 1 \right]^{1/2}~~~~~~~~(\chi < 0)
\eea
where $\eta$ is related to the cosmic time $t$ as $dt(\eta) = b(\eta) d\eta$.

We observe that there is, in fact, no value of $\eta$ for which $b$ vanishes. In other words,
even when one finds a value of $\eta$ for which $t = 0$, then for that value of $\eta$ the scale
factor $b$ remains non-zero. Thus even at the epoch $t = 0$, $b \neq 0$ and $\zeta$ finite ---
there is no big bang singularity.

\section{Conclusion}

In conclusion, we have found that in a spacetime with torsion the axion (dual to the
KR field strength) provides a natural solution to the problem of big bang
singularity present in the FRW model. Such an axion appears naturally in heterotic
string theory. Different kinds of KR-torsion couplings have been considered and we
have shown that for a certain kind of coupling the Universe passes through an
accelerating phase which at least qualitatively explains the recent
experimental findings through the Supernova Ia. 
The possibility of acceleration at different epochs as well as the size of the Universe at the 
initial time in such scenario has been estimated.
We have further shown that the  other scalar massless mode of string theory, namely the dilaton, 
also plays a crucial role to control the expansion rate at various epochs. 
Our work differ from the previous works \cite{coplidsey} significantly. 
Here we explicitly solve for the axion by using it's equation of motion
in different sectors of it's coupling with torsion. 
Although both the phenomena of removal of big bang singularity and
late-time accelerating Universe are found to depend on the dilaton and the KR field
strength, in this work we have focused primarily to extract the effect of the antisymmetric
KR field on these two phenomena. 
This work thus proposes a possible resolution of two of the most important
problems prevailing in the present cosmological scenario.

\vskip .2in
\noindent
{\bf {\Large Acknowledgment}}
\vskip .1in

SS acknowledges the Council of Scientific and Industrial Research, Govt. of India for
providing financial support.

%\end{thebibliography}

\end{document}